\documentclass{elsarticle}

\usepackage{amssymb}

\usepackage[usenames,dvipsnames,svgnames,table]{xcolor}
\usepackage[utf8]{inputenc}
\usepackage{listings}
\usepackage[tight,footnotesize]{subfigure}
\usepackage[normalem]{ulem}
\usepackage[hyphens]{url}

\usepackage{svg}

\journal{...}

\begin{document}

\begin{frontmatter}



\title{Metadata Interpretation Driven Development}


\author[PPgEEC,UFRN]{Júlio~G.S.F.~da~Costa}
\author[GEO,UFRN]{Reinaldo~A.~Petta}
\author[PPgEEC,DCA,UFRN]{Samuel~Xavier-de-Souza\corref{cor1}}
\ead{samuel@dca.ufrn.br}
\address[PPgEEC]{Graduate Program of Electrical and Computer Engineering}
\address[GEO]{Departamento de Geologia}
\address[DCA]{Departamento de Engenharia de Computação e Automação}
\address[UFRN]{Universidade Federal do Rio Grande do Norte, Natal, RN, Brazil, 59078-900}

\begin{abstract}
Despite decades of engineering and scientific research efforts, \emph{separation of concerns} in software development remains not fully achieved. 
The challenge has been to avoid the \emph{crosscutting of concerns} phenomenon, which has no apparent complete solution.
In this paper, we show that business-domain coding plays an even larger role in this challenge. We then introduce a new approach called \emph{Metadata Interpretation Driven Development} (MIDD), which suggests a way to enhance the current way of realizing separation of concerns by  eliminating the need to code functional concerns. 
We
propose to code non-functional concerns as metadata interpreters. This interpretation occurs at run-time and is possible because it assumes the existence of such metadata in artefacts created in previous stages of the process, such as the modelling phase.
We show how this can increase the (re)use of the constructs. Furthermore, we show that a single interpreter,  due to its semantic disconnection from the domain, can simultaneously serve different business domains with no concerns regarding the need to rewrite or refactor code.
Although high-reuse software construction is considered a relatively mature field, changes in the software services scenario demand constant evolution of the actual solutions. The emergence of new software architectures, such as serverless computing, reinforces the need to rethink software construction. This approach is presented as a response to this need. 
\end{abstract}


\end{frontmatter}


\lstset{
language=java, 
basicstyle=\ttfamily\footnotesize, 
keywordstyle=\color{OliveGreen}\bfseries,
identifierstyle=, 
commentstyle=\color{Gray}, 
stringstyle=\color{Magenta}, 
showstringspaces=false} 

\section{Introduction}

{S}{eparation} of Concerns (SoC) is a term coined by Djikstra in 1974 \cite{dijkstra1982role}. The context was the software crisis that the industry and academia had to overcome. The costs associated with the management and development of software became increasingly complex because of the unstructured manner that they were built. The challenge was therefore to implement strategies that could allow the management of projects and the construction of software in a rational, structured and economically viable way.

Regardless of the purpose for which the built software is provided, the structuring principle that arose from this challenge was \emph{componentization}, as a strategy to guide the separation of concerns. Since the advent of componentization, much has been invested to improve the process of building software and its associated artefacts. The result is the emergence of paradigms and methodologies among which we can mention: Structured Development; Object-Oriented Programming (OOP)~\cite{korson1990understanding}; Subject Orientated Programming (SOP)~\cite{harrison1993subject}; Aspects Oriented Programming (AOP)~\cite{kiczales1997aspect}; Multi-Dimensional Separation of Concerns (MSoC)~\cite{tarr1999ndegrees}; Representational State Transfer (REST)~\cite{Richards2006}; Design Patterns~\cite{Gamma1994Design};
Model-Driven Development~\cite{Beydeda2005Model}; Software Product Lines (SPL)~\cite{kang2009applied}; and others~\cite{KITCHENHAM20097}.

Despite investments in all the paradigms mentioned in the previous paragraph, separation of concerns remains relevant. This is due to the ever-increasing complexity of the software management and construction process, which is reinforced today by the centrality of the use of software in quotidian human life~\cite{agnihotri2020systematic, venters2018software, Vale2016TwentyeightYO}.

\subsection{Criticism to Actual Programming Paradigms and Programming Tools} \label{sec:intro_criticism_to_actual_paradigm}
The need for lower costs in software evolution or maintenance makes these topics active areas of research. Rezende~\cite{rezende2005engenharia} and de Vasconcelos et al.~\cite{vasconcelos2017application} estimates that up to 90\% of all costs associated with the life cycle of software is concentrated in activities of maintenance or evolution.

Assuming that OOP is the most relevant paradigm implemented to carry out componentization, research aimed at mitigating the cost of changing requirements can be classified into two axes. The first axis is about the criticism on the form of how the OOP paradigm assimilates and represents real-world objects~\cite{harrison1993subject,kristensen1996conceptual,rayside2000aristotelian}. The second is about how the OOP coding tools implement the notion of a relationship among objects~\cite{pearce2006relationship, bierman2005first, osterbye1999technology, kristensen1994complex}.

On the first axis, the main criticism to OOP is that the coded object only reflects a perspective of how it appears in the real world---perceived as a complex of static (attributes) and dynamic (methods) characteristics. However, an object from the real world can be perceived in the most diverse and sometimes unpredictable ways~\cite{rayside2000aristotelian,harrison1993subject}. This causes difficulties to modify software artefacts that require some customization when circumstances change in its business domain. Studies that have pointed out possible ways to address coding issues related to this problem are SOP~\cite{harrison1993subject}; MDSoC~\cite{tarr1999ndegrees}; HoH~\cite{ernst2003higher}; MAP~\cite{skotiniotis2010modular}; and SPL~\cite{kang2009applied}. Of all these, the most relevant and consistently adopted is SPL, with the main application limit being the most mature business domains in terms of understanding how they exist and work~\cite{rincon2018applies}. The others have not had the same success in terms of adoption.

In the second axis, the goal is to investigate how the OOP coding tools implement relationships. In this regard, Pearce \& Noble~\cite{pearce2006relationship}, Bierman \& Wren~\cite{bierman2005first} and Osterbye~\cite{osterbye1999technology} emphasize that the notion of relationship is very well captured and expressed at the highest levels of abstraction by the currently available tools. However, at the level of implementation, the notion of relationship is not well captured \emph{to facilitate reuse} by the OOP languages. The consequence is coupling among components and therefore difficulties with reuse. 

To address these difficulties, \emph{componentization} and AOP are the main solutions that have emerged as ways to perform decoupling of software concerns. More specifically, AOP promotes the separation between functional and non-functional concerns, as proposed by Kiczales~\cite{kiczales1997aspect}. 

However, employing AOP poses some challenges. One challenge comes from adding a new layer of technology to artefacts, implying, for example, greater concerns about traceability issues among artefacts~\cite{macia2011exploratory, murphy2001software}. Raheman et. al~\cite{RAHEMAN2018562} identified some of these issues and proposed ways to overcome them. Another challenge lays in the requirement for teams with more mature skills, either from the perspective of understanding the software architecture or from the perspective of coding this architecture~\cite{SANTOS201631}.

On another perspective, motivated by the peculiar dynamics of changes in software requirements and the need for controlling costs in the software construction process~\cite{vasconcelos2017application}, a new field of research emerges.
Namely, the quest for tools whose objective is to \emph{intelligently forecast the cost} of the software construction process~\cite{bilgaiyan2017systematic,hryszko2017assessment}. 

All of these issues and the attempts to solve them are symptoms related to a reality in which it is still relevant to research mechanisms that help reduce software building efforts. Such efforts are generally made by adding layers of methodology and/or technologies to processes and artefacts---potentially with cost implications. For the sake of scope, we have not advanced on this aspect of the discussion.


\subsection{Proposition}

Regarding the criticism presented in the previous subsection, we propose an approach for software development named \emph{Metadata Interpretation Driven Development} (MIDD). The objectives are:
\begin{itemize}
  \item remove from the code the need to implement domain business data, i.e. its metadata;
  \item reduce the dependence between non-functional and functional concerns in the data domain, to increase the chance of code reuse; and
  \item decrease pressure for traceability between models and code artefacts.
\end{itemize}

To achieve that, the proposed approach promotes the modelling of data artefacts to the category of first-class artefacts. Here, these artefacts become necessary not only for the understanding of the domain but also and more importantly for the operation of the software system. 
In this approach, the models are consumed during the operation of the software, skipping the implementation of the artefacts that represent them at the code level. 
In summary, \emph{domain metadata interpretation} is the cornerstone that guides the process of building software according to the proposed approach.

\subsection{Paper Organization}
The rest of this paper follows with the Section~\ref{sec:probform} presenting the formulation of two problems tackled by our approach. 
Following, Section~\ref{sec:propappr} contains the description of the proposed approach. 
Then, in Section~\ref{sec:conclusion}, we point out the main contributions of this work and the possible research needs that may unfold from it.

\section{Formulation of the Main Problems}
\label{sec:probform}
In the next two subsections, we present the main challenges related to the construction of software that our approach intends to solve. 
For an illustration of these challenges, we present a software model for the management of car rental offices and, later, another model for book lending management. 

\subsection{Limits Related to the Realization of Relationships among Concerns}
\label{sec:probform1}
\begin{figure}[!thb]
\centering
\includegraphics[height=.46\textheight]{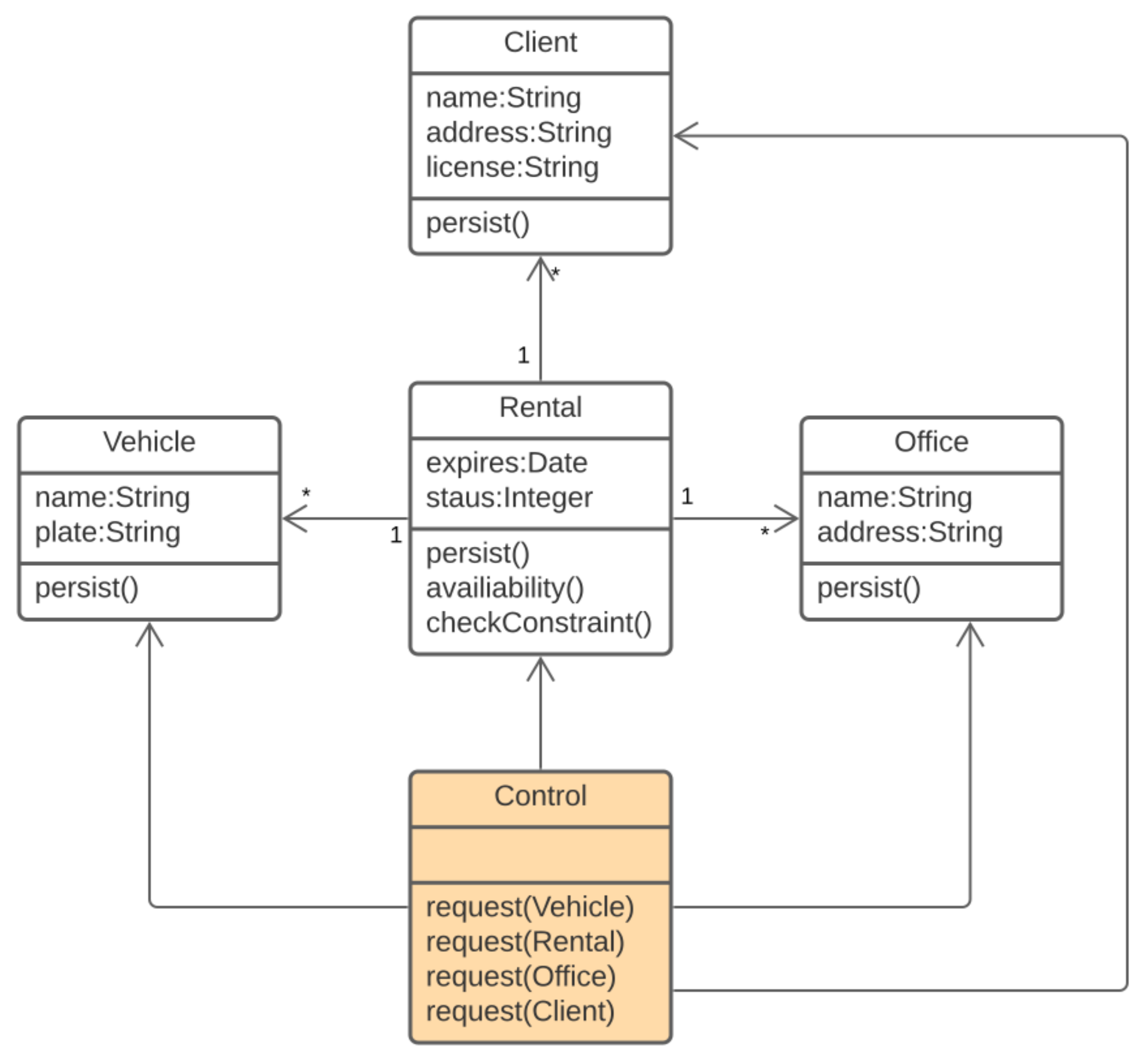}
\caption{Software modelling of a car rental management system.}
\label{fig:carrental}
\end{figure}

Consider the modelling of a car rental system as presented in Fig.~\ref{fig:carrental}. 
It synthesizes the information that the rentals are realized by the establishment of relationships among instances of {\tt Client}, {\tt Vehicle} and {\tt Office}. One business rule required here is that it prohibits renting to customers without a driver's license. 

This first modelling would not be considered good enough given the previously mentioned paradigms and the tools that implement them. 
In this regard, it is important to note that each modelled domain data class has common implementations regarding the business rule and persistence concerns.
There is, at the modelling and coding levels (if modelling is followed), a strong entanglement between the different concerns, in addition to cohesion problems. And these are characteristics that lead to a poorly constructed system from the perspective of separation of interests. This problem is known as \emph{the tyranny of the dominant decomposition}~\cite{kiczales1997aspect}. 
This phenomenon occurs because the data concerns are commonly taken into account more than the others at the moments of analysis and design.

\begin{figure}[!thb]
\centering
\includegraphics[height=.542\textheight]{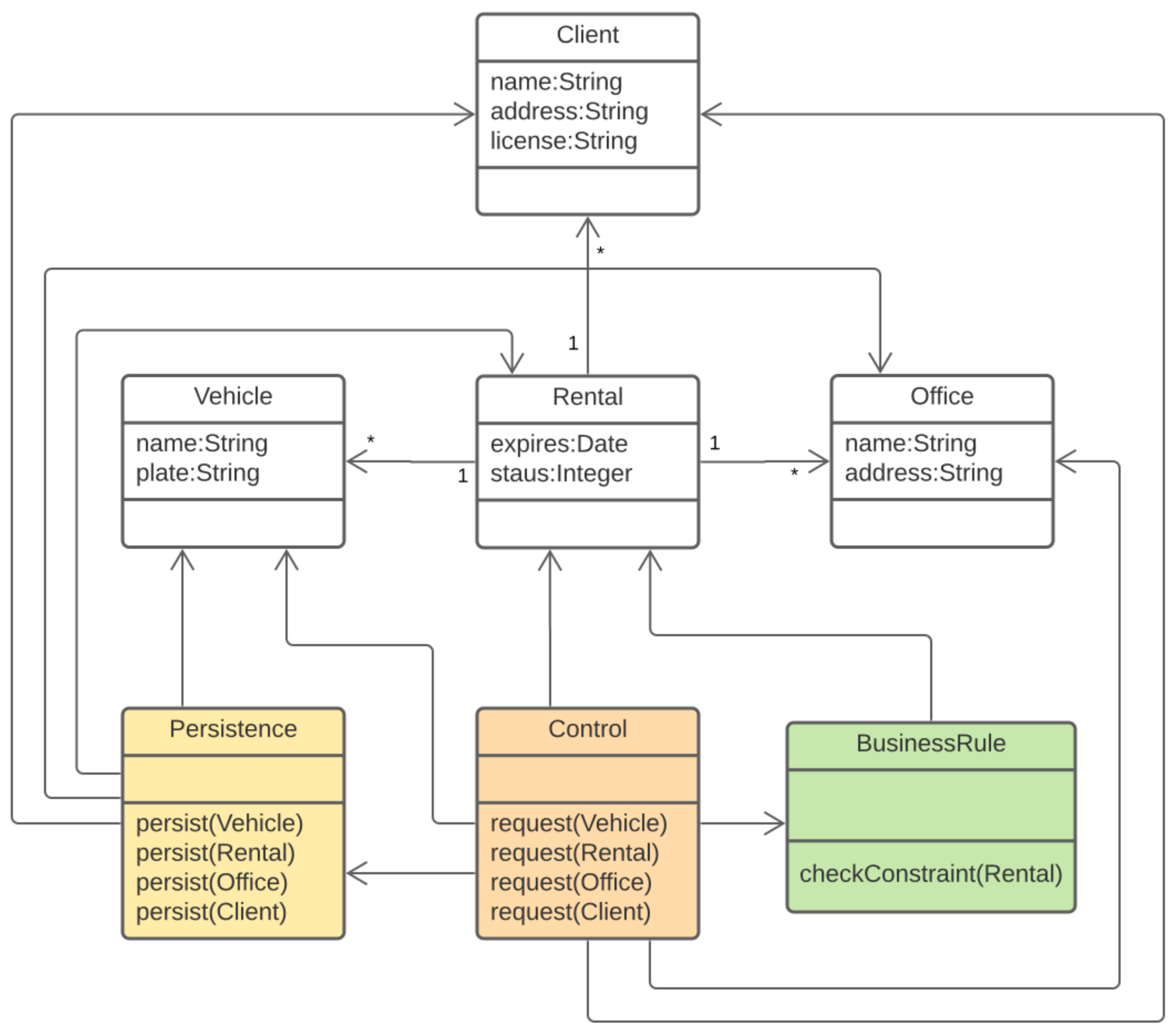}
\caption{Alternative modelling of a car rental management system decomposed on concerns: persistence, business rules and data.}
\label{fig:altcarrental}
\end{figure}

Fig.~\ref{fig:altcarrental} presents a better design for the example of Fig.~\ref{fig:carrental}.
Note that the concerns were refactored to mitigate the degree of spread presented in the first model. 
The non-functional concerns of persistence and business rules were removed from the classes that implement the data concerns ({\tt Vehicle}, {\tt Office}, {\tt Client} and {\tt Rental}) and placed, each one, in classes that implement them in their specificities---{\tt Persistence} and {\tt BusinessRules}. 
This helps to increase the cohesion of the artefacts (like classes, for example) that implement each of the concerns, whatever they may be. But non-functional concerns remain entangled with data concerns, and vice versa. Such non-functional concerns need to be aware of each of the system's domain data classes so that they can compute over them.

Regarding the concern of business rules, treating it as a concern transversal to data as usually done with the persistence concern (or other non-functional concerns such as access control, auditing, transaction) is somewhat controversial in the literature. However, this approach can still be considered an acceptable path.
Some authors acknowledge the weakness that this represents for the principles of object-oriented programming, but argue that business rules should be removed from classes that implement data concerns and placed in their own separate implementation~\cite{dhondt1999using, cibran2003aspects, cibran2007connecting}. In other words, business rules should be considered as a software concern just like the others. This is the path we followed in this work.
%
We decided not to approach the construction of business rules as a target of interest. In the perspective of the proposed approach, the target is to discuss and present a way to guide the construction of the other non-functional concerns. For the examples of domains modelled here in the particular, control and persistence are the target concerns.

On with the new refactoring given in Fig.~\ref{fig:altcarrental}, note that despite the increasing degree of separation of concerns, the scattering and crosscutting of concerns remains at the level of modelling and even coding, see Fig.~\ref{fig:altcarrental}, Fig.~\ref{fig:code1}, Fig.~\ref{fig:codeBR} and Fig.~\ref{fig:code2}. Although the implementation of the persistence concern is now all concentrated in a single new class called {\tt Persistance} (the same is true about the business rules concern), it remains entangled by data concerns---notice, for example, the method signature {\tt persist(Vehicle)} in that class.

\begin{figure}[!thb]
\centering
\begin{lstlisting}
public class Persistence {
  Vehicle read(Vehicle filters) {
    Connection conn = Database.getConnection();
    SQLStatment stt = conn.getStatement();
    stt.getSelectStatementFor(Vehicle).setFilters(filters);
    Vehicle vehicle = stt.execute();
    stt.close();
    conn.close();
    return vehicle;
  }
  
  void insert(Vehicle vehicle) {
    Connection conn = Database.getConnection()
    SQLStatment stt = conn.getStatement();
    stt.getInsertStatementFor(Vehicle).setValues(vehicle);
    stt.execute()
    stt.close();
    conn.close();
  }
  
  Rental read(Rental filters) {
    Connection conn = Database.getConnection();
    SQLStatment stt = conn.getStatement();
    stt.getSelectStatementFor(Rental).setFilters(filters);
    Rental rental = stt.execute();
    stt.close();
    conn.close();
    return rental;
  }
  
  void insert(Rental rental) {
    Connection conn = Database.getConnection()
    SQLStatment stt = conn.getStatement();
    stt.setInsertStatementFor(Rental).setValues(rental);
    stt.execute()
    stt.close();
    conn.close();
  }
  // ...
}
\end{lstlisting}
\caption{A short description in Java about the implementation of the persistence concern in order to exemplify how it crosscut the data concern. In the same class, {\tt Persistence}, there exists instances of {\tt Vehicle} and {\tt Client}, marking the crosscutting of the concerns.}
\label{fig:code1}
\end{figure}

\begin{figure}[!thb]
\centering
\begin{lstlisting}
public class BusinessRules {
  // ...
  Boolean checkConstraints(Rental rental) {
    if (rental.getClient().hasLicense())
      return true;
    else return false;
  }
  // ...
}
\end{lstlisting}
\caption{A short description in Java about the implementation of the business rules concern in order to exemplify how it crosscut the data concern. In the same class, {\tt BusinessRule}, there exists instances of {\tt Rental} and {\tt Client}, marking the crosscutting of the concerns.}
\label{fig:codeBR}
\end{figure}

\begin{figure}[!thb]
\centering
\begin{lstlisting}
class Control {
  //...
  void request(Rental rental, int operation) {
    // ...
    BusinessRules businessRules = new BusinessRules();
    Persistence persistence = new Persistence();
    if (operation == INSERT and businessRules.checkConstraints(rental)) {
      persistence.insert(rental);
    } else if (operation == UPDATE and businessRules.checkConstraints(rental)) {
      persistence.update(rental);
    }
    // ...
  }
  
  void request(Client client, int operation) {
    // ...
  }
  // ...
}
\end{lstlisting}
\caption{A short in Java description about the implementation of a control concern (layer responsible for receiving requests to operate on data states). An example of how it crosscut the concerns: in the same class, {\tt Control}, there exists instances of {\tt Rental}, {\tt Client} and {\tt Persistence}.}
\label{fig:code2}
\end{figure} 

In this regard, the decomposition required to remove all crosscutting of concerns is not possible to be accomplished through the use of mechanisms such as the available object-oriented programming languages~\cite{cibran2003aspects,cibran2007connecting}. Figs.\ref{fig:code1}-\ref{fig:code3} exemplify this. In Fig.\ref{fig:code1}, for example, the implementation of the persistence concern is entangled by the presence of data concerns (e.g. {\tt Vehicle} and {\tt Rental}), as instances of objects of these concerns are declared and used internally to those. The same can be seen in the implementation of the concern for business rules (Fig.~\ref{fig:codeBR}) and control (Fig.~\ref{fig:code2}).

\begin{figure}[!thb]
\centering
\begin{lstlisting}
public class Client {
  public String name;
  public String address;
  public String license;
  public Set<Rental> rentals;
}

public class Vehicle {
  public String model;
  public String plate;
  public Set<Rental> rentals;
}

public class Rental {
  public Integer status;
  public Date expires;
  
  public Client client;
  public Vehicle vehicle;
  public Office Office;
}
\end{lstlisting}
\caption{A short Java implementation of data concerns to exemplify how they crosscut each other, as seen for {\tt Client} and {\tt Rental} and for {\tt Rental}, {\tt Vehicle}, and {\tt Office}.}
\label{fig:code3}
\end{figure}


Moreover, the crosscutting concerns that occur within the dimension of the functional concerns are also an issue~\cite{tarr1999ndegrees}---Fig.~\ref{fig:code3} intends to present this. Note that {\tt Client} has an assignment of {\tt Rental}, and vice-versa; we can say the same for {\tt Vehicle} and {\tt Rental}. The question here is: what if the number of classes in the domain becomes larger? how could we arrange them in data components? If we consider that the domain's data classes may exist in distinct components (for example, a new component), considering each component a specific functional concern, the crosscutting between these components will occur. Therefore, this issue also manifests itself within functional concerns, not just between functional and non-functional issues.

The work of Tarr et al.~\cite{tarr1999ndegrees} contextualize this problem and proposes an approach called Multi-Dimensional Separation of Concerns (MDSoC). 
They highlight the necessity to think about strategies that recognize and decouple distinct data concerns---as distinct data components---to achieve a higher degree of code reuse.

\subsection{Limits Related to the OOP Paradigm}
\label{sec:probform2}
Let us assume that the developers of the Information System (IS) for the previous car rental office have been hired to construct an IS for a school. 
In performing the software requirements analysis, they realize that the part of the school IS concerning the lending of library books has the same operating principle as the IS designed and built for car rental offices. 
However, even if the entire procedure of operations are similar, it is not possible to simply withdraw this operating principle from the car rental system and insert it as part of the school system. 
Changes in the definition of data concerns are necessary and care needs to be taken with possible changes in the other concerns crosscut therein. 
These modifications are necessary because the implementation should be able to adequately represent the context of book lending. 

Two possible paths arise here to solve this problem in the most efficient and scalable way. 
In the first, the concepts in common remain and the divergent ones are removed from the original implementation, creating a new implementation for the domain. 
The second design option is to refactor the business domain of book lending and car rental in a more abstract domain, such as rent of goods and services, which is what they have in common. 

\begin{figure}[!thb]
\centering
\includegraphics[height=.425\textheight]{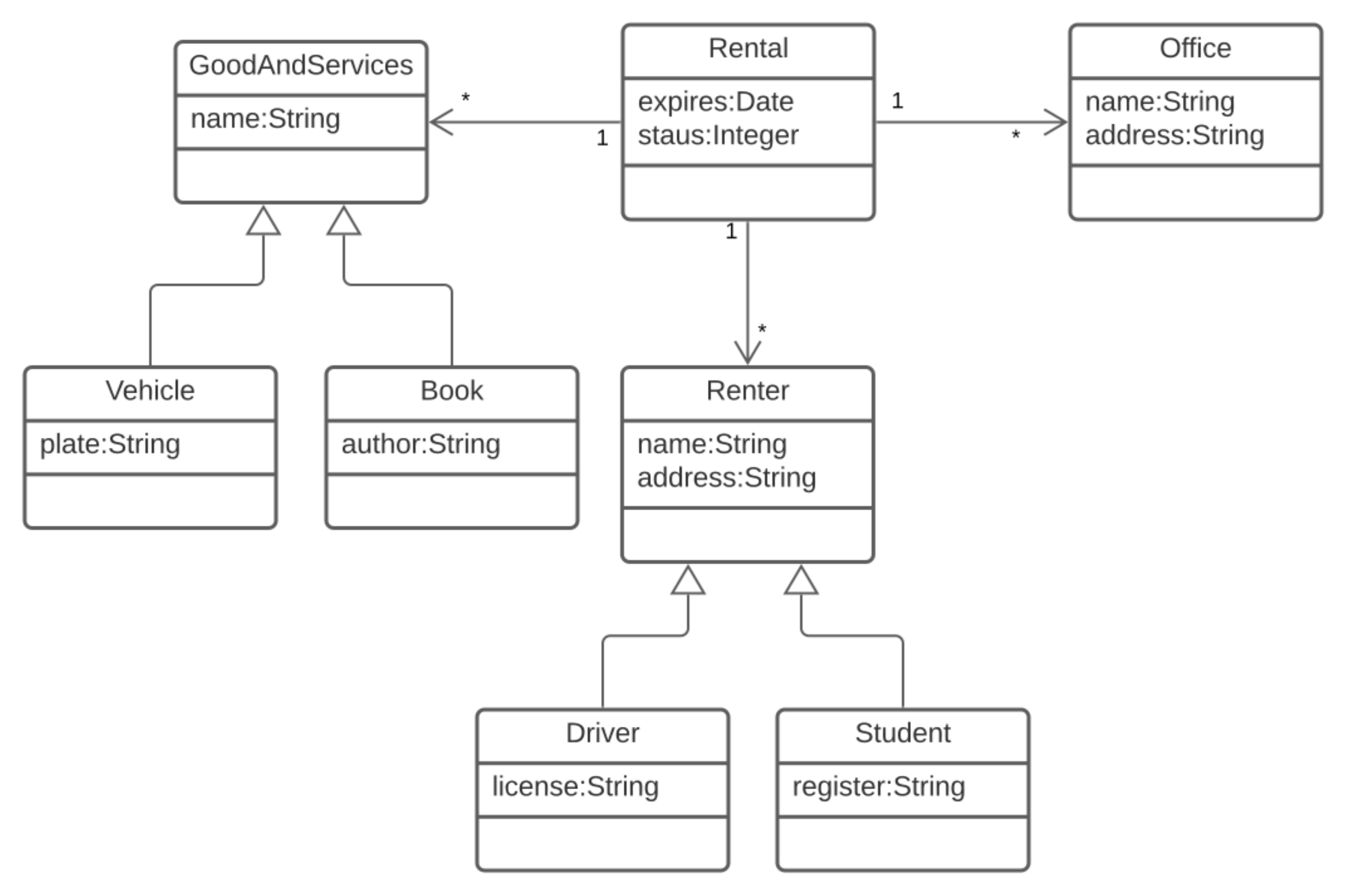}
\caption{A refactorization of the data models for rental office and library in a data model for rent of goods and services that abstracts both business domains.}
\label{fig:factoreddomain}    
\end{figure}

The latter path is the commonly preferred design because it seeks to preserve what in principle is considered proper among several possible instances of the rent-of-goods-and-services business domain. 
However, it increases the complexity of the domain model and, as already mentioned, synchronization between the representations of the modelling and coding layer remains a challenge. Nevertheless, this seems likely to be the best path. A new design taking this path could be the one seen in Fig.~\ref{fig:factoreddomain}.
Additional to significantly altering the original car rental model, the most relevant difficulty here is the need to make significant changes also at the code level.
When new data concerns arise, new relationships among these concerns are demanded and so, consequently, the difficulties of maintaining relationships discussed in subsection~\ref{sec:probform1} also arise. 
In this sense, a lot of effort must be made to keep \emph{the implementation} always consistent with \emph{the modelling} for both systems. 

On the other hand, someone could claim that the professionals involved in building the car rental IS---analysts and designers---have not been able to predict other possible uses for the software structure. 
However, methods that leverage the ability to make this prediction is an area that needs more research.
The so-called \emph{oracle hypothesis} is a key element in the success of reuse and domain engineering~\cite{frakes2005software}.
It requires from analysts the ability to predict the future, or the ability to contemplate, in the processes of analysis, the maximum predictability about how the system construction can be applied to different contexts, which is not always possible. 
Nevertheless, there are cases where this requirement for a predictive domain analysis can be used with some satisfactory degree of efficiency. 
This is the case, for example, of business domains that are already mature and stable. 
In these cases, it is recommended the use of the Software Product Lines (SPL) paradigm, whose process methodology takes into account refactorization of domains to achieve efficiency in terms of reuse through prediction of possible uses of the constructed software components~\cite{kang2009applied}.

Given this issue that concerns the reuse of artefacts---implying the need for a good understanding of business domains, a good representation of it in models and their coding---the next section presents the proposed Metadata Interpretation Driven Development approach, as a form to improve the reuse of artefacts by suppressing the domain data representation at the code level.
For the two IS models discussed here, we present how this new approach circumvents the circumstanced problems to present better possibilities of reuse the constructed artefacts.

\section{Metadata Interpretation Driven Development}
\label{sec:propappr}
Northover et al.~\cite{northover2008towards}, reinforced by de Vasconcelos~\cite{vasconcelos2017application}, state that the challenge of building software is to be able to accommodate design the changes coming from the most varied sources of pressure---whether they are technological, economic, social, or even ideological---at whatever moments they arise.
However, even if we recognize this pressure for modifying the definition of things---names and meanings---in a business domain, we still need to capture these things, requiring some level of stability of their names and meanings to properly compute them, as pointed out by Rayside and Campbell~\cite{rayside2000aristotelian}. 

And here we could identify a dilemma: to conciliate the fluidity of the meanings on real-world things with the inflexibility of their representation on the computing-world data structures. 
This is the cornerstone for a series of reflections on the philosophical aspects of software engineering and for the contributions of this work.

\subsection{A Philosophical Motivation}
Giguette~\cite{giguette2006building}, Northover et al.~\cite{northover2008towards} and Rayside and Campbell~\cite{rayside2000aristotelian} argue about the necessity of a more philosophical investigation of the \emph{reality phenomena} to make professionals more capable and efficient to construct software systems. 
They state categorically that the OOP paradigm is based on Plato's philosophy, more specifically, all of them relates \emph{object orientation} to the notion of \emph{ideal form} developed in Plato's ideas. In this regard, Giguette~\cite{giguette2006building} writes: 
\begin{quotation}
``According to Plato, each real-world object is an approximation of its \emph{ideal form}. 
A particular tree is an instance of the Tree form, which defines the characteristics all trees have in common. 
Plato might say that a Java class is similar to a platonic form because it generalizes all possible class instances``.
\end{quotation}

Rayside and Campbell~\cite{rayside2000aristotelian} attempt to understand the paradigm of building object-oriented software by the understanding of the Aristotelian philosophy. 
For the authors, both Aristotle and Plato were engaged in solving a central problem of Greek philosophy: how to reconcile the intelligibility of the real world with the fact that things are changing (both its static and dynamic aspects)? 
The same type of question that an analyst or a software developer needs to make before enterprising in the construction of software.

Aristotle's answer regarding the intelligibility of things is in the existence of a universal character in them, something that transcends the uniqueness of an individual and reaches everyone in a given scope. 
Aristotle uses the concepts of genus (a name for a class) and species (a name for an object) to argue about the intelligibility of things despite the changes. 
For him, a species (or an object) represents the intelligible form of singularities classified from it, whereas the genus (or the class) represents the more abstract and transcendent intelligible form. 
This is exactly the principle in the object-oriented programming paradigm: the classification of things in terms of classes and objects, or terms of genus and species.

Confronting what is said in the works of Rayside and Campbell~\cite{rayside2000aristotelian}, Giguette~\cite{giguette2006building} and Northover et al.~\cite{northover2008towards} with the actual forms of software development, we perceive that the spreading of the coupling among software concerns is difficult to be removed at the level of code, this is because all artefacts (models and codes) are constructed as representations of things in the real world. 

Given what is said by these authors, we suggest that the difficulty in removing the coupling between the various concerns, which constitute the classes of objects in software construction, is related to the fact that all constructs are representations of the real world and, as the real world is always changing, these representations are pressured to be updated---the problem of achieving \emph{the ideal form} of objects. 

That is, this seems much more a problem about the changing nature of real-world objects (or this understanding) than a problem about coding.
Therefore, the reason for the root of software evolution costs remains clear: concepts change, so codes change, and this always comes at a cost. 
In the proposed Metadata Interpretation Driven Development, operating at the meta-metadata level rather than at metadata level in coding mitigates the effect caused by the problem of crosscutting software concerns.
In this new approach, there is no longer the need to implement, at the code level, the metadata of data concerns identified in a given business domain. By doing that, we mitigate the spread of coupling between data concerns with each other (discussed at the end of subsection ~\ref{sec:probform1}) and other concerns (non-functional concerns, for example).



In the following subsection, we present how the models discussed in sections~\ref{sec:probform1} and~\ref{sec:probform2} can be implemented according to the MIDD approach by shifting the implementation to the meta-metadata level.

\subsection{Realizing MIDD}
The MIDD approach demand the construction of data models. 
For example, they can be constructed in terms of a modelling language like Unified modelling Language (UML), or even in terms of another language that matters to the particular business domain. 
Data models of the domain become essential artefacts to the operation of the software system because the information about the data concerns are not translated to the code level. 
This information is now interpreted at runtime by other non-functional concerns.

The approach waives the implementation of all data classes, e.g. {\tt Student}, {\tt Vehicle} and {\tt Book}. On the contrary, such code, now constructed as a metadata interpreter (according to the desired interests---persistence, access control, transaction, etc.) can be applied to any business domain in which the construction of codes as metadata interpreters is applied. 

Therefore, all other software concerns must be implemented from the perspective of interpretation of the domain metadata. 
This approach deviates from the conventional process of implementing/coding software concerns. 
The non-functional concerns must deal with the metadata interpretation so that they can realize the contracted non-functional expectations of the given software system.
Note that our approach does not suggest opposition to using the OOP paradigm as a way to realize non-functional concerns, nor as a way to model real-world objects. Instead, OOP can be seen as defining the way the interpreter should recognize and operate on the domain's data instance.
However, we emphasize that this is done without additional layers of tools in the construction process.
\begin{figure}[!thb]
\centering
\includegraphics[height=.675\textheight]{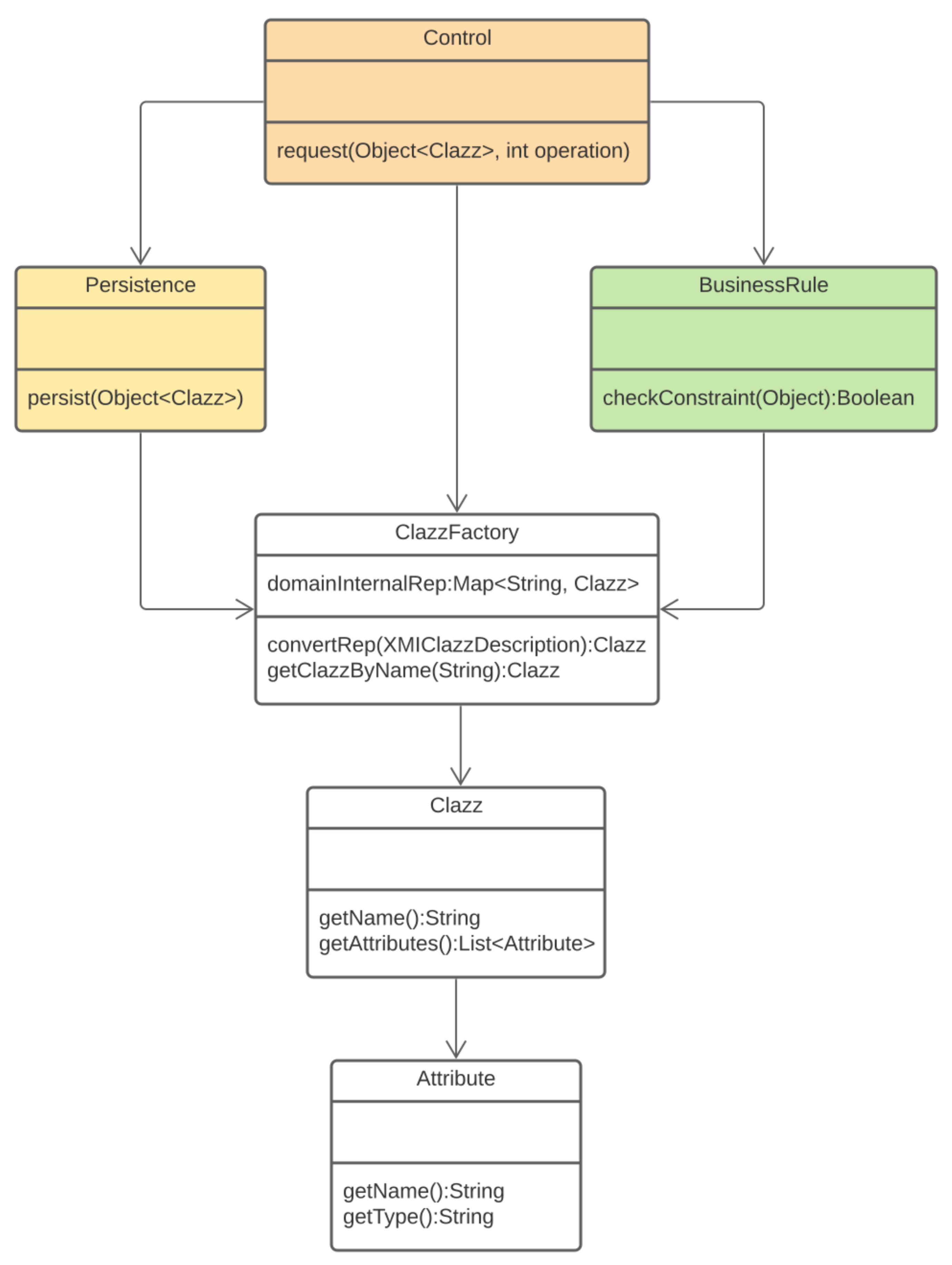}
\caption{Details of the proposed approach. Data concerns are removed from the code level.}
\label{fig:proposal}
\end{figure} 

In the following subsection, we examine what happens with the concerns of control and persistence at the code level, comparing their form with the form presented in subsection~\ref{sec:probform1}.

\subsection{Instaciating MIDD}

The process of instantiating metadata in a meta-metadata structure is performed in two actions: first, the data classes are captured in terms of their metadata, for which we want computation, from a model document (a UML class diagram, for example) during software initialization; second, the metadata are associated with their internal representation in the system at runtime. This internal representation is equivalent to what it represents in the domain model, a document made during the software modelling phase. 
The existence of this representation avoids a new recovery of the model at each request to operate on the same instance of domain data.
Fig.~\ref{fig:code5} and Fig.~\ref{fig:code6} illustrate the implementation of this process.
\begin{figure}[!tbh]
\centering
\begin{lstlisting}
class Clazz {
  //XMIClazzDescription: possibly part of an UML Classes Diagram
  Clazz(XMIClazzDescription xmiClazzDesc) {
    //...
  }
  String getName() {
    //...
  }
  Set<Attribute> getAttributes() {
    //...
  }
}

class Attribute {
  //XMIAttributeDescription: possibly derived from XMIClazzDescription
  Attribute(XMIAttributeDescription xmiAttributeDesc) {
    //...
  }
  String getNane() {
    //...
  }
  String getType() {
    //...
  }
}
\end{lstlisting}
\caption{The metadata that describe what is a Clazz and Attribute. Note that to instantiate an object of this type it is necessary to pass to them a semantic value that will describe them in the context of the operations demanded from them.}
\label{fig:code5}
\end{figure} 

\begin{figure}[!tbh]
\centering
\begin{lstlisting}
class ClazzFactory {
  // See Clazz definition above
  Map<String, Clazz> domainInternalRep;
  // XMI: possibly an UML Classes Diagram
  ClazzFactory(XMIModelDescription xmiModelDesc) {
    //...
    for (XMIClazzDescription xmiClazzDesc : xmiModelDesc.getClazzes()) {
      domainInternalRep.put(xmiClazzDesc.getName(), convertRep(xmiClazzDesc));
    }
    //...
  }
  //...
  Clazz convertRep(XMIClazzDescription xmiClazzDescription) {
    Clazz clazz;
    // do representation convertion, from external (XMI) to internal (Clazz)
    return clazz;
  }
  //...
  Clazz getClazzByName(String name) {
    return domainInternalRep.get(name);
  }
  //...
}
\end{lstlisting}
\caption{ {\tt ClazzFactory} is the mechanism that realizes the metadata that describe the data concern in order to be interpreted by the non-functional concerns---view, persistence, business rules, etc.}
\label{fig:code6}
\end{figure} 

Fig.~\ref{fig:code5} presents a possible implementation of the meta-metadata for wrapping metadata structures, i.e. for getting a representation of the data domain, at runtime, for any class regarding any data object that exists in the system's business domain---a way to remove the implementation of metadata from the code level. 
Furthermore, another important aspect is the mechanism needed to instantiate the classes described by Fig.~\ref{fig:code5} that specify the domain data objects instantiated by the system. The mechanism that starts and builds the internal representation of the domain model is the class {\tt ClazzFactory}, as shown in Fig.~\ref{fig:code6}. 

The purpose of {\tt ClazzFactory} is to obtain the set of metadata from the domain model (see class {\tt XMIModelDescription} as a parameter for the initialization of {\tt ClazzFactory} in Fig.~\ref{fig:code6}), organize them class by class (see class {\tt XMIClazzDescription}) in an internal representation (see the method {\tt convertRep} and the data structure {\tt domainInternalRep}) and then make it available to be queried regarding the specification of any class in the domain for any software concern (for example, persistence, transaction, visualization, etc).

So far, we have presented a mechanism for retrieving metadata, building their internal representation, and accessing them to reference any data requested to be processed. Now, through Fig.~\ref{fig:codeBRAlt}, Fig.~\ref{fig:code7} and Fig.~\ref{fig:code8}, we present the issues related to the scattering of crosscutting between data and business rules concerns, between data and persistence concerns, and between data and control concerns.

\begin{figure}[!thb]
\centering
\begin{lstlisting}
public class BusinessRules {
  // ...
  Boolean checkConstraints(Object object) {
    // Making a cast to get an rental object, from object
    Rental rental = (Rental) object;
    if (rental.getClient().hasLicense())
      return true;
    else return false;
  }
  // ...
}
\end{lstlisting}
\caption{A short description in Java of the implementation of the business rules concern to exemplify how it crosscuts the data concern despite the MIDD usage approach. The class {\tt BusinessRule} contain an instance of {\tt Rental}.}
\label{fig:codeBRAlt}
\end{figure}

\begin{figure}[!thb]
\centering
\begin{lstlisting}
class Control {
  //...
  //object: represents a data class instance containing a set of
  //        tuples attribute name/value, like this:
  //  { "clazz":"Vehicle", "model":"Volkswagen Golf", "plateNumber":"WE231CK" }
  //...
  request(Object object, int operation) {
    // ...
    if (operation == CREATE and BusinessRules.checkConstraints(object)) {
      Persistence.insert(object);
    } else if (operation == UPDATE and BusinessRules.checkConstraints(object)) {
      Persistence.update(object);
    }
    // ...
  }
  // ...
}
\end{lstlisting}
\caption{The Control concern implemented as a metadata interpreter in Java.}
\label{fig:code7}
\end{figure} 
\begin{figure}[!thb]
\centering
\begin{lstlisting}
class Persistence {
  //...
  //object: represents a data class instance containing a set of
  //        tuples attribute name/value, that serve for filtering:
  //  { "clazz":"Vehicle",  "model":"Volkswagen Golf" }
  //...
  List<Object> read(Object object) {
    Clazz clazz = ClazzFactory.getClazzByName(object.get("clazz"));
    Connection conn = Database.getConnection();
    Statement sttm = conn.getStatement();
    List<Object> objects = sttm.executeSelectStatementFor(clazz, object);
    sttm.close();
    conn.close();
    return objects;
  }
  
  //...
  //object: represents a data class instance containing a set of
  //        attribute name/value, like this:
  //  { "clazz":"Vehicle", "model":"Volkswagen Golf", "plateNumber":"MS23C09" }
  //...
  void insert(Object object) {
    Clazz clazz = ClazzFactory.getClazzByName(object.get("clazz"));
    Connection conn = Database.getConnection();
    Statement sttm = conn.getStatement();
    sttm.executeInsertStatementFor(clazz, object)
    sttm.close();
    conn.close();
  }
  //...
}
\end{lstlisting}
\caption{ The Persistence concern implemented as a metadata interpreter in Java.}
\label{fig:code8}
\end{figure} 

Instead of coding the {\tt Control} class in a way that favours the spreading of concerns as we show in Fig.~\ref{fig:code2} (see the method signatures {\tt request(rental, operation)}, {\tt request(client, operation)}), in the MIDD approach, we code only one method, {\tt request(object, operation)}, for any object of any domain data class---Fig.~\ref{fig:code7} and Fig.~\ref{fig:code8} attempt to elucidate this. 
The reader is invited to compare these figures with Fig.~\ref{fig:code1} and Fig.~\ref{fig:code2} to establish a first comparison between the two approaches. 

Regarding the separation of concerns between data and business rules, as informed in subsection \ref{sec:probform1}, the implementation follows as previously given in Fig.~\ref{fig:codeBR}. 
The only difference lays in the signature of the {\tt checkConstraint} method, in the class {\tt BusinessRules}, in Fig.~\ref{fig:codeBRAlt}. Or, in another way, the crosscutting of these concerns is postponed from the method signature to the method body.

Metadata interpretation arises as a new software concern.
However, by coding non-functional concerns as metadata interpreters, we can avoid coding data concerns, whatever is the business domain. 
To the extent of this paper, the above sentence is true at least for the control and persistent concerns.

The reduction of issues regarding traceability across concerns is another benefit of the approach.
These issues arise, above all, with the need to update classes of data concerns due to misunderstandings, or due to the evolution of the business domain. 
Updates need to be performed, in general, everywhere in the code where the specific data concern occurs.

Since the non-functional concerns no longer have to deal with data objects from a specific business domain, signing their definition in code, the services of non-functional concerns can be implemented in a more reusable way. 
Thus, the proposed approach contributes to improving the degree of code reuse. 
For example, the {\tt Control} and {\tt Persistence} classes can now be reused for many other data classes from any business domain, including those for which it has not been foreseen.

Building the non-functional concerns as interpreters works fine for the modelling proposed in Fig.~\ref{fig:altcarrental}, as well for the modelling in Fig.~\ref{fig:factoreddomain}, or even for a third model that represents a variation of the goods-and-services domain, or relative to any other business domain. Note that we do not promote a new formalism, such as a new language for representing concerns. 
Redefining the way we codify the non-functional concerns is enough to promote an increase in reuse.

Lastly, the data models become here prevalent for the proper functioning of the system.
The data class hierarchy must represent well the data concepts to leverage better reuse of the captured data intelligence of a business domain---however, this data representation does not need to exist at the code level. 
In a sense, MIDD requires solid modelling of the data business domain, which is in itself beneficial for the maintenance and evolution of the information system.

\subsection{Implementing Persistence Concern as an Interpreter}

The persistence concern implementation is intended to prepare and organize the data in a certain format so that they can be stored, usually, in long-term memory. In this sense, we assume the existence of a relational database for the needs demanded by this work. Since at the software system level we are dealing with an object-oriented data representation, the persistence concern implementation must perform an object-relational mapping here.

In the context of MIDD, this mapping takes place at the time of calling the method {\tt sttm.executeInsertStatementFor(clazz, object)} or when calling the method {\tt sttm.executeSelectStatementFor(clazz, object)}, see Fig.~\ref{fig:code8}; and are implemented in the class {\tt Statement} in Fig.~\ref{fig:orm}. 
\begin{figure}[!thb]
\centering
\begin{lstlisting}
class Statement extends SQLStatement {
  //...
  //object: represents a data class instance containing a set of
  //        tuples attribute name/value, that serve for filtering:
  //  { "clazz":"Vehicle",  "model":"Volkswagen Golf" }
  //...
  List<Object> executeSelectStatementFor(Clazz clazz, Object object) {
    StringBuilder predicate = new StringBuilder();
    
    object.keys().forEach(key -> {
      if (clazz.getAttributes().contains(key)) {
        predicate.concat(key).concat("=").concat(object.get(key)).concat(" AND ");
      }
    });
    
    List<Object> objects = new List<Object>
    
    List<Row> rows = super.execute("SELECT * FROM {table} WHERE {predicate}"
      .replace("{table}", clazz.getName())
      .replace("{predicate}", predicate.removeLast(" AND ")));
    
    rows.forEach(row -> {
      Object object = new Object();
      clazz.getAttributes().forEach(attribute -> {
        object.put(attribute.getName(),row.getValueByColumnName(attribute.getName()));
      });
      objects.put(object);
    });
    
    return objects;
  }
  
  //...
  //object: represents a data class instance containing a set of
  //        attribute name/value, like this:
  //  { "clazz":"Vehicle", "model":"Volkswagen Golf", "plate":"MS23C09" }
  //...
  Statement executeInsertStatementFor(Clazz clazz, Object object) {
    StringBuilder cols = new StringBuilder();
    StringBuilder values = new StringBuilder();
    
    clazz.getAttributes().forEach(attribute -> {
        cols.concat(attribute.getName()).concat(", ");
        values.concat(object.get(attribute.getName())).concat(", ");
    });
    
    super.execute("INSERT INTO {table} ({cols}) VALUES ({values})"
      .replace("{table}", clazz.getName())
      .replace("{cols}", cols.removeLast(", "))
      .replace("{values}", values.removeLast(", ")));
  }
}
\end{lstlisting}
\caption{ The Statement class implemented as an object-relational mapping to MIDD in Java.}
\label{fig:orm}
\end{figure} 

In the MIDD, for the persistence concern, the object-relational mapping takes place at runtime, at the instant operations on data are demanded, and without data concerns being represented at the code level. On the contrary, as far as we know from our research, knowing two of the most relevant tools\footnote{For example Hibernate (http://hibernate.org/) for the Java language, and the Doctrine (https://www.doctrine-project.org/projects/orm.html) for PHP language.} to perform this type of mapping, it happens at \emph{programming} and \emph{startup time}, requiring the implementation of data concerns at the code level.

In contrast to what is done by other approaches (cited in the previous paragraph), that deal with performing interpretation, we want to emphasize the MIDD contribution for this persistence concern implementation: the metadata interpretation, as seen in the control concern and here in more detail must occur \emph{at runtime}---precisely to avoid the coupling between non-functional and functional concerns at \emph{programming} or \emph{startup} time. In this sense, if we want the same type of separation of concerns achieved for the implementation between the persistence and data concerns presented in this section, all other non-functional concerns (such as vision, access control, transaction, etc.), we must take into account the proposition of the MIDD approach.  

\clearpage
\section{Final Remarks}
\label{sec:conclusion}
The collect effort spent over a few decades to improve processes and tools to satisfy the demand for good software components---as cohesive and uncoupled as possible with high rates of reuse---is still relevant and new approaches that offer gains in terms of increasing rates of reuse of software artefacts continue to be desired.

We showed that a relevant barrier to achieving separation of concerns comes from the fact that what is coded, in terms of data concerns, are representations of the definition of data in a software business domain. The need to represent the business domain at the code level implies that for every change in what is represented there will be a change in the code representation wherever it appears---and that always entails some cost. In addition, and in order to mitigate this effort, predicting all domain configurations, offering a model and code factorization for all foreseen possibilities, is something that requires great skills from professionals and cannot be applied to all business domains.

Taking these issues into consideration, we propose a way to carry out the separation of concerns called Metadata Interpretation Driven Development (MIDD). Centrally, our approach suggests removing, as much as possible, the functional concerns of the software system implementation.
We propose that domain data metadata is not implemented at the coding level.
Instead, they must be interpreted at runtime.
This implies that the software needs to be developed with the ability to interpret the set of metadata that describes the data domain for which it operates at runtime. 
We offer an implementation path for the proposed MIDD approach that results in the creation of a metadata interpreter. 
In principle, this interpreter can be employed to act as a persistence concern service for data from any business domain. 
We recognize, however, that the same cannot be stated of the business rule concern. 
We have not addressed business rules concerns through the MIDD approach, whose implementation remains tied to the other software development approaches.

We reinforce, to promote a broader understanding of everything that has been exposed, that a proof-of-concept implementation of a MIDD interpreter is available in the following repository: \url{https://gitlab.com/lappsufrn/MIDD}. Along with the code made in the Java language, there are instructions available to instantiate and test the MIDD interpreter.

Finally, we understand that there is a lot to be investigated w.r.t. the proposed approach. 
For example, what impacts does this approach have on activities such as testing and deployment? How much overhead will there be for the system due to the need to interpret metadata in addition to performing the concern operation that is implemented as an interpreter? What are the challenges of taking this approach to other concerns such as transaction, security, view, etc? Which are the gains and limitations that such an approach has compared to current strategies and architectures that are employed to realize separation of concerns? The answers to such questions are the subject of future work as it escapes the introductory scope of this work. 

\section{Acknowledgments}
\label{}
The authors would like to thank André~H.~Siqueira and Zalkind~L.D.~Rocha for the inspirational discussions.\\
Funding: This work was supported by Coordenação de Aperfeiçoamento de Pessoal de Nível Superior—Brazil (CAPES)—Finance Code 001.



 \bibliographystyle{elsarticle-num} 
 \bibliography{refs_}





\end{document}